\documentclass[prb,nofootinbib,twocolumn,superscriptaddress]{revtex4} 

\usepackage{graphicx}
\usepackage{dcolumn}
\usepackage{bm}
\usepackage{threeparttable}
\usepackage{times}
\usepackage{mathptmx}
\usepackage{lscape}
\usepackage{natbib}
\usepackage{amsmath}
\usepackage{amssymb}
\usepackage{braket}
\usepackage{comment}
\usepackage{color}

\def\degree{\kern-.2em\r{}\kern-.3em}

\begin{document}


\title{  Efficient, Systematic Estimation of Alloy Free Energy from Special Microscopic States  }
   
\author{Ryogo MIyake}
\affiliation{
Department of Materials Science and Engineering,  Kyoto University, Sakyo, Kyoto 606-8501, Japan\\
}%
      
\author{Subaru Sugie}
\affiliation{
Department of Materials Science and Engineering,  Kyoto University, Sakyo, Kyoto 606-8501, Japan\\
}%

\author{Koretaka Yuge}
\affiliation{
Department of Materials Science and Engineering,  Kyoto University, Sakyo, Kyoto 606-8501, Japan\\
}%

\begin{abstract}
{ 
For classical discrete systems under constant composition typically refferred to substitutional alloys, we propose calculation method of Helmholtz free energy based on a set of special microscopic states. The advantage of the method is that configuration of the special states are essentially independent of energy and temperature, and they depend only on underlying lattice: The special states can be known a priori without any thermodynamic information, enabling systematic prediction of free energy for multicomponent alloys. 
We confirm that by comparing to conventional thermodynamic simulation, information about the special states provide reasonable predictive power above order-disorder and phase-separating transition temperature for alloys with many-body (up to 3-body) interactions. 
  }
\end{abstract}


\maketitle

\section{Introduction}
In order to achieve suitable alloy design, Helmholtz free energy should be cental and fundamental information to construct e.g., their phase diagrams. 
For classical discrete system under constant composition typically refferred to multicomponent alloys, free energy can be given by
\begin{eqnarray}
\label{eq:f}
F &=& -\beta^{-1} \ln Z \nonumber \\
Z &=& \sum_{d} \exp \left( -\beta U^{\left( d \right)} \right)
\end{eqnarray}
where $\beta$ denotes inverse temperature, Z partition function, and summation is taken over possible microscopic states on configuration space (i.e., possible configurations). 
 It is clear from Eq.~\eqref{eq:f} that with increase of system size, the number of possible configurations exponentially increases, making  the direct estimation of $F$ practically intractable. Therefore, various calculation techniques have been amply developed, including entropic sampling with Monte Carlo (MC) simulation the cluster variation method and thermodynamic integration with Frenkel method.\cite{mc1,mc2,mc3,wl}
 Basically, the existing calculation method explore a set of configurations that dominantly contributes to estimation of $F$, where these configurations should in principle depend on temperature and energy through Eq.~\eqref{eq:f}. 
 
 Another important point for $F$ calculation is accurate estimation of the potential energy $E$ for each configuration, where first-principles (FP) total- energy calculation should be employed. However, direct estimation of $E$ for possible configurations requires huge amount of computational cost. One of the most effective approaches to overcome the problem is to combine FP calculation with generalized Ising model (GIM),\cite{gim} where energy for any given configuration $d$ can be exactly expressed as
\begin{eqnarray}
U^{\left( d \right)} = \sum_{r=1}^{f} \Braket{U|q_{r}} q_{r}^{\left( d \right)},
\end{eqnarray}
where summation is taken over structural degree of freedoms (SDFs), $\left\{ q_{r} \right\}$ denotes a set of complete basis functions of GIM, and $\Braket{\quad|\quad}$ represent inner product, i.e., trace over possible configurations. Since $\left\{ q_{r}^{\left( d \right)} \right\}$, corresponding to multisite correlations, can be determined from configuration and $\left\{ \Braket{U|q_{r}} \right\}$ is independent of configuration, we should know optimal values of $\left\{ \Braket{U|q_{r}} \right\}$ for given alloy system through FP calculation, typically requires at least several hundreds of total-energy calculations for representative configrations. 
 Although the procedure of how to select a set of the representative configurations has been amply improved through several theoretical studies, it still remains difficult to apply the GIM-based procedure with FP calculation when we \textit{systematically} change conditions including the constituent elements for multicomponent systems: This is because a set of representative configurations should depend on alloy system, which cannot be known a priori without any thermodynamic information. 

In order to overcome the above practical problem, here we propose a new free energy calculation method so that a set of representative configuration is \textit{independent} of energy and temperature: This can be achieved by rewriting semi-grand canonical average of composition in terms of multivariate momenets based on GIM basis functions, and applying suitable dimentionality reduction.  
We confirm that the proposed method can reasonablly predict composition-dependence of free energy above the transition temperature, by comparing with conventional thermodynamic simulation. 
Details of the concept, the derivation, and related discussion are shown in the following.

\section{Derivation and Applications}
Let us first briefly explain GIM basis functions used in the present work, which describes configuration and corresponding energy. 
We here focus on binary A-B system up to 3-body correlation, where occupation of given lattice site $i$ is specified by spin variable $\sigma_{i}=1$ for A and $\sigma_{i}=-1$ for B. Then, $k$-th basis function for configuration $d$ is given by
\begin{eqnarray}
q_{k}^{\left( d \right)} = \Braket{ \prod_{i\in S_{k}} \sigma_{i} }_{d},
\end{eqnarray}
where $\Braket{\quad }_{d}$ denotes taking linear average for configuration $d$, and the product is taken over a set of symmetry-equivalent figure $S_{k}$ of $k$. Hereinafter, withoug lack of generality, we describe configuration $d$ measured from center of gravity for configurational density of states (CDOS), which is the density of states for possible configurations before applying many-body interaction to the system.

Under this preparation, we next consider semi-grand canonical average of point cluster $q_{1}$ (i.e., on-site correlation):
\begin{eqnarray}
\label{eq:y}
\Braket{q_{1}}_{Y} = Y^{-1} \sum_{d} q_{1}^{\left( d \right)} \exp \left\{ -\beta \left( U^{\left( d \right)} - \Delta\mu N_{A}^{\left( d \right)} \right) \right\},
\end{eqnarray}
where $Y$ denotes semi-grand partition function defined as
\begin{eqnarray}
Y = \sum_{d} \exp \left\{ -\beta \left( U^{\left( d \right)} - \Delta\mu N_{A}^{\left( d \right)} \right) \right\}.
\end{eqnarray}
Here, $\Delta\mu = \mu_{A} - \mu_{B}$ denotes difference in chemical potential per atom for constituents A and B, and $N_{A}$ denotes number of A in the system. 
When we know the relationships between chemical potential difference and semi-grand canonical average of $q_{1}$ given by
Eq.~\eqref{eq:y}, we can estimate corresponding system free energy through the conventional thermodynamic relation:
\begin{eqnarray}
F\left( T,x_{A} \right) = F\left( T,x_{A}=0 \right) + N \int_{x'_{A}=0}^{x'_{A}=x_{A}} \Delta\mu d x'_{A},
\end{eqnarray}
where $x_{A}^{\left( d \right)}=N_{A}^{\left( d \right)}/N = \left( q_{1}^{\left( d \right)} + 1 \right)/2$.
In order to figure out a set of special configurations independent of energy and temperature, which can characterize behavior of $F$, we first rewrite Eq.~\eqref{eq:y} as 
\begin{eqnarray}
\Braket{q_{1}}_{Y} = \frac{2}{N}\frac{\partial}{\partial m } \ln \left\{ \sum_{N_{A}=0}^{N} \sum_{c} \exp\left( \sum_{i\neq 1}h_{i}\cdot q_{i}^{\left( c \right)} \right) e^{mN_{A}} \right\} -1,
\end{eqnarray}
where 
\begin{eqnarray}
m=\beta\left( \Delta\mu - \frac{2V_{1}}{N} \right) = \beta\left( \Delta\mu - 2v_{1} \right)
\end{eqnarray}
with $V_{i} = \Braket{U|q_{i}} = Nv_{i}$, $h_{i}=-\beta V_{i}$, and summation for $c$ denotes possible configuration on given $N_{A}$. Then we perform Maclaurin series expansion in terms of $\left\{ h_{i} \right\}$ up to the second order and employ exact formulation of multivariate moments under constant composition,\cite{mom} leading to (see Appendix)
\begin{widetext}
\begin{eqnarray}
\label{eq:exp}
\Braket{q_{1}}_{Y} &=& \tanh \frac{m}{2} - 2\beta\left[ \left( \sum_{i} v_{2i} \right) \frac{\partial}{\partial m }  \left( \tanh\frac{m}{2} \right)^{2} + \left( \sum_{j} v_{3j}\right) \frac{\partial}{\partial m }\left( \tanh\frac{m}{2} \right)^{3}   \right] \nonumber \\
&+& \beta^{2} \left[  \left\{  \sum_{i}\sum_{j}\frac{\delta_{ij}^{\left[ 2,3 \right]}}{D_{i} }v_{2i}v_{3j} \right\} \frac{\partial}{\partial m } \left( \tanh\frac{m}{2}  \right)     \right]  \nonumber \\
&+& \beta^{2} \left[  \left\{  \sum_{i}\sum_{j}\left( 4 - \frac{2^{\delta_{ij}}}{Di} \right) v_{2i}v_{2j} + \sum_{i}\sum_{j} \left( f_{ij}^{\left[ 3,3 \right]} - \frac{3\delta_{ij}}{T_{i} } \right) v_{3i}v_{3j}  \right\} \frac{\partial}{\partial m } \left( \tanh\frac{m}{2} \right)^{2}   \right] \nonumber \\
&+& \beta^{2} \left[  \left\{  \sum_{i}\sum_{j}\left( 6-\frac{2\delta_{ij}^{\left[ 2,3 \right]}}{D_{i}} \right) v_{2i}v_{3j}   \right\} \frac{\partial}{\partial m }\left( \tanh\frac{m}{2} \right)^{3}  \right] \nonumber \\
&+& \beta^{2}  \left[  \left\{  \sum_{i}\sum_{j} \left( 9-2f_{ij}^{\left[ 3,3 \right]} + \frac{3\delta_{ij}}{T_{i}} \right)v_{3i}v_{3j} - \sum_{i}\sum_{j} \left( 4-\frac{\delta_{ij}}{D_{i}} \right) v_{2i}v_{2j}      \right\}  \frac{\partial}{\partial m }\left( \tanh\frac{m}{2} \right)^{4}   \right] \nonumber \\
&-&\beta^{2} \left[ \left\{ \sum_{i}\sum_{j}\left( 6 - \frac{\delta_{ij}^{\left[ 2,3 \right]}}{D_{i}} \right) v_{2i}v_{3j} \right\} \frac{\partial}{\partial m }\left( \tanh\frac{m}{2} \right)^{5}   \right] \nonumber \\
&-&\beta^{2} \left[ \left\{ \sum_{i}\sum_{j} \left( 9 - f_{ij}^{\left[ 3,3 \right]} + \frac{\delta_{ij}}{T_{i}} \right) v_{3i}v_{3j} \right\}  \frac{\partial}{\partial m } \left( \tanh\frac{m}{2} \right)^{6}    \right],
\end{eqnarray}
\end{widetext}
where $v_{kn}$ denotes $n$-th figure of $k$-body $\left( k=2,3 \right)$ figure, $\delta_{ij}$ Kronecker delta, $D_{i}$ ($T_{j}$) the number of $i$-th pair ($j$-th triplet) figure per site, and $\delta_{ij}^{\left\{ 2,3 \right\}}$ number of $i$-th pair included in a single $j$-th triplet figure. 
We here emphasize the technically important point in Eq.~\eqref{eq:exp} is that we exclude variable of $\beta\Delta\mu$ from the series expansion so that only \textit{special} configuration-based information (as seen later) directly relates to the series expansion, which is a desirable condition: If we also include $\beta\Delta\mu$ as variable for the expansion, we confirm that undesirable divergence for $\Delta\mu - x$ relationships around equiatomic composition always appears even well above the transition temperature.
When we individually define index for constituent pair in $i$-th ($j$-th) triplet as $2_{i1}$, $2_{i2}$ and  $2_{i3}$ ($2_{j1}$, $2_{j2}$ and  $2_{j3}$), $f_{ij}^{\left\{ 3,3 \right\}}$ is given by
\begin{eqnarray}
f_{ij}^{\left\{ 3,3 \right\}} = \sum_{m=1}^{3}\sum_{n=1}^{3}\frac{\delta_{i_{m}j_{n}}}{D_{i_{m}}}.
\end{eqnarray}

From the first-order term in Eq.~\eqref{eq:exp}, we can clearly see that when 2- and 3-body correlations for two \textit{special} configurations $Q_{g1}$ and $Q_{g2}$ are given by
\begin{eqnarray}
\label{eq:ps1}
Q_{g1} = r_{1}\cdot \left( q_{2_{1}}=1,\cdots,q_{2_{s}}=1,q_{3_{1}}=0,\cdots,q_{3_{t}}=0 \right) \nonumber \\
Q_{g2} = r_{2}\cdot (\left\{) q_{2_{1}}=0,\cdots,q_{2_{s}}=0,q_{3_{1}}=1,\cdots,q_{3_{t}}=1 \right) 
\end{eqnarray}
with arbitrary nonzero real number of $r_{1}$ and $r_{2}$, energy for each configuration satisfies
\begin{eqnarray}
\sum_{i}v_{2i} = r_{1}^{-1} E_{g1},\quad \sum_{j}v_{2j} = r_{2}^{-1} E_{g2}.
\end{eqnarray}
From Eq.~\eqref{eq:ps1}, since configuration $Q_{g1}$ and $Q_{g2}$ can be known a priori without any thermodynamic information, the first-order term of Eq.~\eqref{eq:exp} can be evaluated from energy of the two \textit{special} configurations that are common for any constituents. 

For the second-order terms in Eq.~\eqref{eq:exp}, we first define the following vector:
\begin{eqnarray}
\mathbf{v} = \left( v_{2_{1}},\cdots, v_{2_{s}},v_{3_{1}},\cdots, v_{3_{t}} \right).
\end{eqnarray}
Then Eq.~\eqref{eq:exp} can be rewritten as the following form:
\begin{eqnarray}
\label{eq:re}
\Braket{q_{1}}_{Y} &=& \tanh \frac{m}{2} - 2\beta \left[  \sum_{k=1}^{2} \left\{ r_{k}^{-1}E_{gk} \frac{\partial}{\partial m }\left( \tanh\frac{m}{2} \right)^{k+1} \right\}  \right] \nonumber \\
&+& \beta^{2} \sum_{n=1}^{6} \left[ b_{n} \cdot\mathbf{v}^{\textrm{T}}\mathbf{A}_{n}\mathbf{v} \left\{ \frac{\partial}{\partial m } \left( \tanh\frac{m}{2} \right)^{n} \right\}  \right] ,
\end{eqnarray}
where $b_{n}$ takse +1 for $n=1,2,3,4$ and -1 for $n=5,6$, and $\mathbf{A}_{n}$ is $\left( s+t \right)\times\left( s+t \right)$ real symmetric matrix whose elements corresponds to each coefficient of $v_{ki}v_{lj}$ in Eq.~\eqref{eq:exp}. 
To obtain a set of special configurations for the second-order term, we perform singular value decomposition (SVD) for $\mathbf{A}_{n}$, namely
\begin{eqnarray}
\mathbf{A}_{n} = \sum_{i=1}^{a} \lambda_{i}^{\left( n \right)} \mathbf{u}_{i}^{\left( n \right)} \mathbf{u}_{i}^{\left( n \right)\textrm{T}}
\end{eqnarray}
where $\lambda_{i}^{\left( n \right)}$ and $\mathbf{u}_{i}^{\left( n \right)}$ respectively corresponds to the $i$-th singular value and sigular vector for $\mathbf{A}_{n}$, and $a$ denotes effective rank of $\mathbf{A}_{n}$. 
Therefore, we can specify a set of spesial configuration $Q_{ fi}^{\left( n \right)} $, satisfying
\begin{eqnarray}
\label{eq:ps2}
Q_{fi}^{\left( n \right)} = r_{i}^{\left( n \right)} \cdot \mathbf{u}_{i}^{\left( n \right)}
\end{eqnarray}
with arbitrary nonzero real number of $r_{i}^{\left( n \right)}$, Eq.~\eqref{eq:re} can be further rewritten by using energy for the additional special configurations:
\begin{eqnarray}
\label{eq:fin}
\Braket{q_{1}}_{Y} &=& \tanh \frac{m}{2} - 2\beta \left[  \sum_{k=1}^{2} \left\{ r_{k}^{-1}E_{gk} \frac{\partial}{\partial m }\left( \tanh\frac{m}{2} \right)^{k+1} \right\}  \right] \nonumber \\
&+& \beta^{2} \sum_{n=1}^{6} \left[ b_{n} \cdot \left\{ \sum_{i=1}^{a}\left(  \frac{ E_{fi}^{\left( n \right)}  }{r_{i}^{\left( n \right)} } \right) ^{2}      \right\}    \left\{ \frac{\partial}{\partial m } \left( \tanh\frac{m}{2} \right)^{n} \right\}  \right] .\nonumber \\
\quad
\end{eqnarray}
From Eq.~\eqref{eq:ps2}, since configurations $\left\{ Q_{fi}^{\left( n \right)} \right\}$ can also be known a priori without any thermodynamic information, Eq.~\eqref{eq:fin} can be evaluated from energy of the two and additional \textit{special} configurations that are common for any constituents. 
Eq.~\eqref{eq:fin} can be effectively expressed by a few number of special configuration particularly for low-rank $\mathbf{A}_{n}$, and we confirm that this can typically hold for representative lattices including fcc and bcc. 

\begin{figure}[h]
\begin{center}
\includegraphics[width=0.83\linewidth]{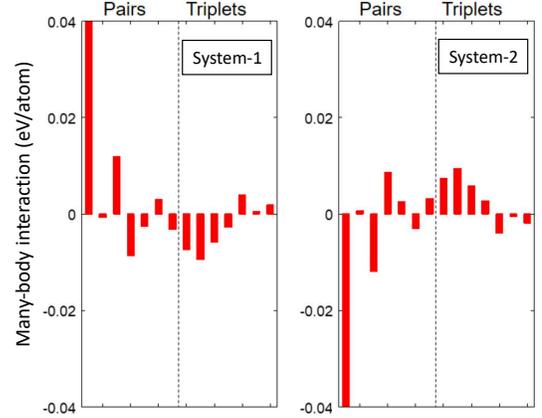}
\caption{ Many-body interaction up to 3-body correlation for two prepared system on fcc lattice.}
\label{fig:eci}
\end{center}
\end{figure}

We finally demonstrate applicability of the proposed approach for estimation of free energy, preparing two artificial fcc-based equiatomic system with many-body interactions shown in Fig.~\ref{fig:eci}, respectively exhibit order structure and phase separation for the ground state. 
\begin{figure}[h]
\begin{center}
\includegraphics[width=1.0\linewidth]{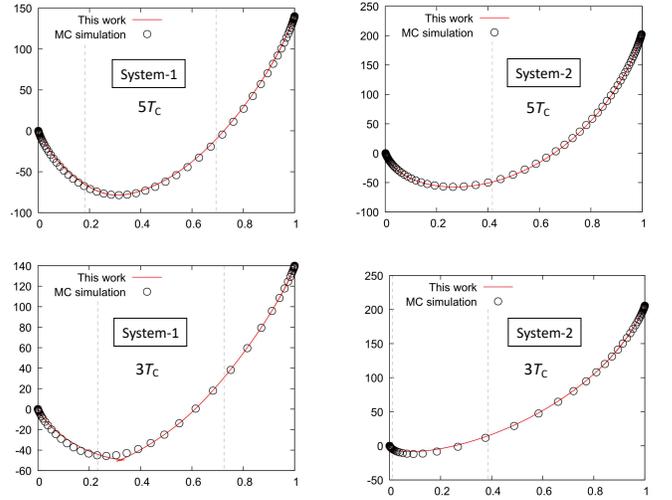}
\caption{Composition-dependence of free energy for two systems at $T=5T_{\textrm{C}}$ and $3T_{\textrm{C}}$.}
\label{fig:f}
\end{center}
\end{figure}

The resultant composition-dependence of free energy at $T=5T_{\textrm{C}}, 3T_{\textrm{C}}, T_{\textrm{C}}$ are shown in Fig.~\ref{fig:f}, compared with the results obtained from conventional thermodynamic simulation with Metroplis algorism. 
The results clearly show that above transition temperature of $T\ge 3T_{\textrm{C}}$, the proposed method provide reasonable agreement with thermodynamic simulation, where the proposed methed employs information only about energy of special configurations common for the two systems and does not required explicit information about many-body interaction or not require sampling across configuration space, while the conventional one should require both information. 
The significant difference in free energy around transition temperature mainly comes from lack of higher-order moment information, which should be addressed in our future study. 

\section{Conclusions}
We propose a new estimation method of Helmholtz free energy from information about energy of special configuration, which can be known a priori without any thermodynamic information, and can be determined only from lattice geometry. The proposed method demonstrates reasonable agreement in free energy prediction compared with conventional thermodynamic simulation above the transition temperature, indicating that the proposed method can effectively provide systematic prediction of alloy phase stability for various combination of constituets.

\section{Acknowledgement}
This work was supported by Grant-in-Aids for Scientific Research on Innovative Areas on High Entropy Alloys through the grant number JP18H05453 and a Grant-in-Aid for Scientific Research (16K06704) from the MEXT of Japan, Research Grant from Hitachi Metals$\cdot$Materials Science Foundation.


\begin{thebibliography}{9}
\bibitem{mc1} N. Metropolis, A. W. Rosenbluth, M. N. Rosenbluth, A. H. Tellerand, and E. Teller, J. Chem. Phys. \textbf{21}, 1087 (1953).
\bibitem{mc2} A. M. Ferrenberg and R. H. Swendsen, Phys. Rev. Lett. \textbf{63}, 1195 (1989). 
\bibitem{mc3} J. Lee, Phys. Rev. Lett. 71, 211 (1993).
\bibitem{wl} F. Wang and D.P. Landau, Phys. Rev. Lett. \textbf{86}, 2050 (2001).
\bibitem{gim} J. M. Sanchez, F. Ducastelle, and D. Gratias, Physica \textbf{A28}, 334 (1984).
\bibitem{mom} S. Ohta and K. Yuge, J. Phys. Soc. Jpn. \textbf{88}, 034802 (2019).
\end{thebibliography}
\end{document}